\documentclass[prl,preprint,showpacs,superscriptaddress]{revtex4}
\usepackage{graphicx}
\usepackage{indentfirst}
\usepackage{epsfig}
\usepackage{subfigure}
\usepackage{amsmath}
\usepackage{amssymb}
\def\be{\begin{eqnarray}}
\def\ee{\end{eqnarray}}

\begin{document}
\title{Dipole emission from a real cavity in a random medium.\\
      Fluorescence in a homogeneous medium II}
\author{M. Donaire}
\email{manuel.donaire@uam.es}
\address{Departamento de F\'{\i}sica de la Materia Condensada, Universidad Aut\'{o}noma de
Madrid, E-28049 Madrid, Spain.}
\begin{abstract} In this paper we
derive a general expression for the emission rate of an emitter which is 
placed at the center of a real cavity drilled in a generic random medium. We apply our formalism to the computation of
 the decay rate of a single atom which seats within a small molecule embedded in a continuous medium.
\end{abstract}
 \maketitle
\begin{figure}[h]
\includegraphics[height=6.2cm,width=12.cm,clip]{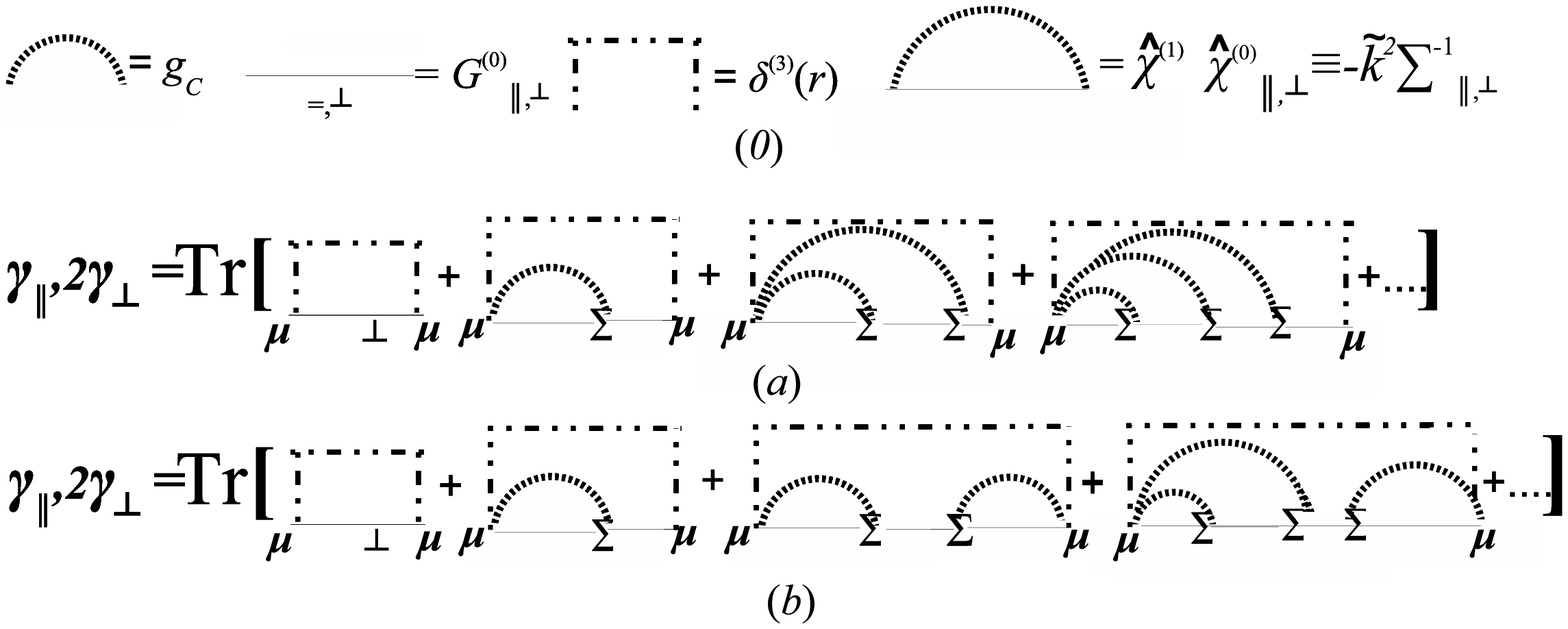}
\caption{($0$) Feynman's rules. ($a,b$) Diagrammatic representation
of the transverse and longitudinal $\gamma$-factors,
according to Eqs.(\ref{gammaanper},\ref{gammaanpar}) ($a$) and Eqs.(\ref{gammabnper},\ref{gammabnpar}) ($b$) in the macroscopic cavity limit.}\label{FigIV01}
\end{figure}
\begin{figure}[h]
\includegraphics[height=12.8cm,width=17.5cm,clip]{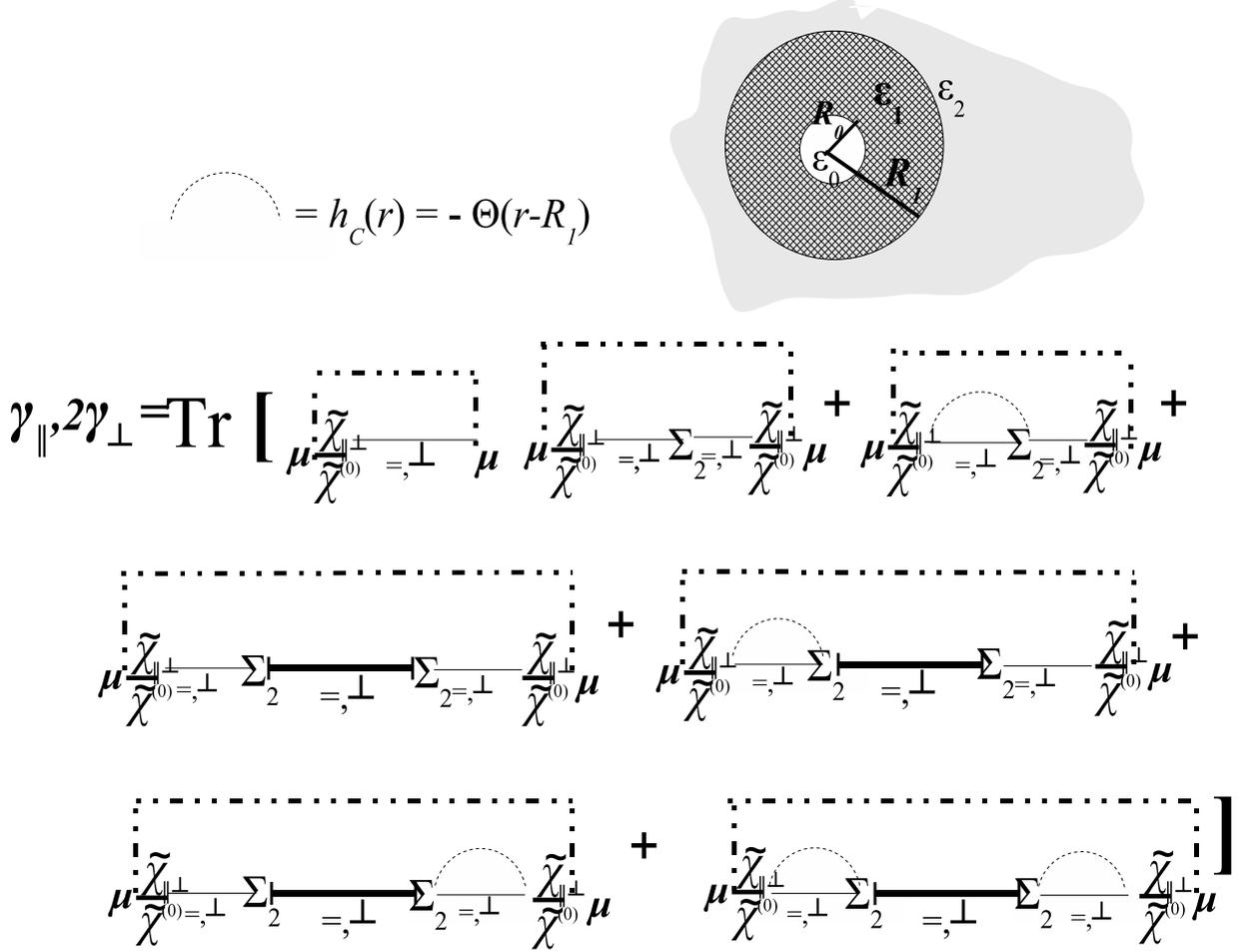}
\caption{Diagrammatic representation of the $\gamma$-factors for the case of a point emitter seated in a small molecule embedded in a continuous medium of permitivity $\epsilon_{2}$. The propagating transverse component is that in Eq.(\ref{firstp}).}\label{FigIV03}
\end{figure}

\indent In a previous paper \cite{MeIII} we computed the spontaneous emission rate of a fluorescent particle seated
at the center of a spherical cavity of radius $R$ drilled in a continuous
medium for the case that $R$ satisfies $R\ll\lambda$, $R\gg\xi$, with $\lambda$ being the wave length of emission and $\xi$ the correlation length of the medium constituents. We referred to the former inequality as \emph{small cavity limit}. For the case that both inequalities hold, the whole system cavity-host-medium can be treated as a continuum and a homogenous bulk propagator can be computed in good approximation. In \cite{MeIII} we used the formal expression obtained in \cite{MeI} for a real cavity scenario in the small cavity limit. In the present paper, we will relax the conditions about the continuity of the system. We will still consider that the host medium is continuous as seen from the emitter beyond $r>R$. That is, $R\gg\xi$. We will refer to this inequality as \emph{macroscopic cavity limit}. However, the freedom in the choice of the ratio $\lambda/\xi$ implies that spatial dispersion in the dielectric constant is allowed. Secondarily,  the lack of constraint in the choice of the ratio $\lambda/R$  implies that the emitted wave does
not only see the cavity twice in the processes of emission and back-reaction, but also from any scattering event within the host medium. Propagation rather depends continuously on the cavity and a translation-invariant
bulk propagator in the medium surrounding the emitter cannot be strictly computed.\\
\indent Our first goal is to derive a recurrent formula which accounts for the whole series of radiative corrections over the bare polarizability function of a point emitter. On the other hand, we would like it to be comparable with the exact formula found in \cite{MeI} for a virtual cavity scenario. Following \cite{MeIII} we will consider a point
 polarizable emitter. Its polarizability function $\alpha(\tilde{k})$ gets renormalized as a
result of iterative self-polarization processes,
\begin{equation}\label{Lorentzpol}
\alpha(\tilde{k})=\alpha_{0}[1+\frac{1}{3}\alpha_{0}\tilde{k}^{2}\Re{\{2\gamma_{\perp}^{(0)}\}}
+\frac{1}{3}\alpha_{0}\tilde{k}^{2}(2\gamma_{\perp}+\gamma_{\parallel})]^{-1}.
\end{equation}
In the above equation, the $\gamma$-factors encode the radiative corrections over the bare polarizability \cite{MeI}. The term
$\frac{1}{3}\alpha_{0}\tilde{k}^{2}\Re{\{2\gamma_{\perp}^{(0)}\}}$ has been introduced to account for the internal resonance
of the emitter in vacuum. As shown in \cite{deVriesRMP}, it plays the role of a regulator of the intrinsic ultraviolet divergence in $G^{(0)}_{\perp}$. That is, $\Re{\{2\gamma_{\perp}^{(0)}\}}=\frac{-3}{k_{0}^{2}\alpha_{0}}$, where $\alpha_{0}$ is the corresponding real electrostatic polarizability and $k_{0}$ is the resonance wave number in vacuum. Following \cite{deVriesPRL}, by parametrizing
Eq.(\ref{Lorentzpol}) in  a Lorentzian (L) form, $\alpha_{L}(\tilde{k})=\alpha'_{0}k_{res}^{2}[k_{res}^{2}-\tilde{k}^{2}-i\Gamma^{\alpha}\tilde{k}^{3}/(c k_{res}^{2})]^{-1}$,
we can identify the decay rate of the emitter in the host medium as
\begin{eqnarray}\label{Gamares}
\Gamma^{\alpha}&=&-\frac{c}{3}\alpha'_{0}\tilde{k}^{3}\Im{\{2\gamma_{\perp}+\gamma_{\parallel}\}}|_{\tilde{k}=k_{res}}
\nonumber\\&=&-\Gamma_{0}\frac{2\pi}{k_{0}^{2}}\tilde{k}\Im{\{2\gamma_{\perp}+\gamma_{\parallel}\}}|_{\tilde{k}=k_{res}},
\end{eqnarray}
where $k_{res}$ is a real non-negative root of the equation
\begin{equation}\label{kres}
(\tilde{k}/k_{0})^{2}-1=\frac{1}{3}\alpha_{0}\tilde{k}^{2}\Re{\{2\gamma_{\perp}+\gamma_{\parallel}\}}|_{\tilde{k}=k_{res}},
\end{equation}
$\alpha'_{0}=\alpha_{0}(k_{0}/k_{res})^{2}$ is the renormalized electrostatic polarizability and $\Gamma_{0}=c\alpha_{0}k_{0}^{4}/6\pi$ is the in-vacuum emission rate. As argued in \cite{deVriesPRL}, consistency with Fermi's golden
rule requires $\alpha_{0}=\frac{2|\mu|^{2}}{\epsilon_{0}\hbar c k_{0}}$, $\mu$ being the transition amplitude between
two atomic levels in vacuum.\\
In the case that the emitter itself is a host scatterer, the emitter occupies a virtual cavity and the factors $\gamma_{\perp,\parallel}$ are given by
\begin{eqnarray}
2\gamma_{\perp}&=&
\int\frac{\textrm{d}^{3}k}{(2\pi)^{3}}\Bigl[\frac{2\chi_{\perp}(k)}{\rho\alpha}G_{\perp}(k)\Bigr]-2\Re{\{\gamma^{(0)}_{\perp}\}},
\label{LDOSIper}\\
\gamma_{\parallel}&=&\int\frac{\textrm{d}^{3}k}{(2\pi)^{3}}\:\Bigl[\frac{\chi_{\parallel}(k)}{\rho\alpha}G_{\parallel}(k)-\frac{1}{\tilde{k}^{2}}\Bigr].\label{LDOSIparal}
\end{eqnarray}
In the above formulae, $\bar{\chi}(k)$ is the effective susceptibility and $G_{\perp,\parallel}$ are the transverse and longitudinal components of the bulk propagator respectively.\\
\indent In the present case, the emitter seats in a real empty cavity of radius $R$. Therefore, it does not correlate to the host medium as any host particle would do. Assuming $R\gg\xi$,
the complete series of diagrams is that in Fig.\ref{FigIV01}, where we use the diagrammatic conventions
employed in \cite{MeI} in the macroscopic cavity limit.\\
\indent The first thing to notice is that the condition $R\gg\xi$ allows for simplifications. Let $g(r)$ be the  two-point correlation function of the emitter with the host scatterers,
\begin{equation}
g(r)=1+h_{C}(r),
\end{equation}
where $h_{C}(r)$ is the irreducible piece. Because the cavity is macroscopical, it distorts in general the homogenous distribution of host scatterers in the surrounding medium. Thus, $h_{C}(r)$ must be provided. For the moment, let us think of it as a step function which accounts for the cavity exclusion volume, $h_{C}(r)\approx-\Theta(r-R)$. Its Fourier transform reads
\begin{equation}
g(k)=(2\pi)^{3}\delta^{(3)}(\vec{k}) + h_{C}(k).
\end{equation}
Because $h_{C}(r)$ has support in $r\leq R$, it is therefore expected that its Fourier transform gets support for $k\lesssim 1/R$. On the other hand, in case that spatial dispersion in $\bar{\epsilon}(k)$ be relevant, such a dispersion must be of the order of $1/\xi$. Therefore, any convolution of $g(|\vec{k}-\vec{k}'|)$ with the self-energy operator $\bar{\Sigma}(k')$ and any generic function $f(k')$ can be approximated by
\begin{equation}\label{laprox}
\int\frac{\textrm{d}^{3}k}{(2\pi)^{3}}g(|\vec{k}'-\vec{k}|)\bar{\Sigma}(k)\cdot\bar{f}(k)\approx
\bar{\Sigma}(k^{'})\cdot\int\frac{\textrm{d}^{3}k}{(2\pi)^{3}}g(|\vec{k}'-\vec{k}|)\bar{f}(k).
\end{equation}
$\bar{\Sigma}(k)$ is a tensor made of one-particle-irreducible (1PI) multiple-scattering processes which, for a statistically homogeneous host medium, can be decomposed as
$\bar{\Sigma}(k)=\Sigma_{\perp}(k)(1-\frac{\vec{k}\cdot\vec{k}}{kk})+\Sigma_{\parallel}(k)\frac{\vec{k}\cdot\vec{k}}{kk}$.
In Eq.(\ref{laprox}), the difference between $\vec{k}$ and  $\vec{k}'$ is negligible in comparison to $1/\xi$ in the range of momenta where $h_{C}(|\vec{k}-\vec{k}'|)$
 takes nearly constant value and $\bar{\Sigma}(k)$ is not zero. In turn, that implies that the correlation of the emitter with any 1PI (multiple)scattering process can be approximated by the correlation of the emitter to any of the scatterers involved in such a process. In particular, things get mathematically simpler if correlations connect either the entrance or the exit scatterer in the corresponding 1PI diagrams. This allows to write the series of radiative correction  as two apparently different expansions, $(a)$ and $(b)$, as depicted in Figs.\ref{FigIV01}$(a,b)$. The corresponding series read
\begin{eqnarray}
2\gamma_{\perp}&=&\sum_{n=0}^{\infty}2\gamma_{\perp}^{a(n)},\qquad\qquad \gamma_{\parallel}=\sum_{n=0}^{\infty}\gamma_{\parallel}^{a(n)},\label{seriesa}\\
2\gamma_{\perp}&=&2\gamma^{(0)}_{\perp}+\sum_{n=1}^{\infty}2\gamma_{\perp}^{b(n)},\quad \gamma_{\parallel}=\gamma^{(0)}_{\parallel}+\sum_{n=1}^{\infty}\gamma_{\parallel}^{b(n)}\label{seriesb}.
\end{eqnarray}
In the above series, the $n^{th}$ term of each one reads
\begin{eqnarray}
2\gamma^{a(n)}_{\perp}&=&2\int\frac{\textrm{d}^{3}k}{(2\pi)^{3}}\hat{\chi}^{(n)}_{\perp}(k)\Sigma_{\perp}(k)G_{\perp}^{(0)}(k),
\label{gammaanper}\\
\gamma^{a(n)}_{\parallel}&=&\int\frac{\textrm{d}^{3}k}{(2\pi)^{3}}\hat{\chi}^{(n)}_{\parallel}(k)\Sigma_{\parallel}(k)G_{\parallel}^{(0)}(k),
\label{gammaanpar}
\end{eqnarray}
and
\begin{eqnarray}
2\gamma^{b(n)}_{\perp}=2\int\frac{\textrm{d}^{3}k}{(2\pi)^{3}}\hat{\chi}^{(n-1)}_{\perp}(k)\Sigma_{\perp}(k)G_{\perp}^{(0)}(k)
\Sigma_{\perp}(k)\hat{\chi}^{(1)}_{\perp}(k),
\label{gammabnper}\\
\gamma^{b(n)}_{\parallel}=\int\frac{\textrm{d}^{3}k}{(2\pi)^{3}}\hat{\chi}^{(n-1)}_{\parallel}(k)\Sigma_{\parallel}(k)G_{\parallel}^{(0)}(k)
\Sigma_{\parallel}(k)\hat{\chi}^{(1)}_{\parallel}(k).
\label{gammabnpar}
\end{eqnarray}
The recurrent formulae for the \emph{partial pseudo-susceptibilities} $\hat{\chi}^{(n)}_{\perp,\parallel}$ --the reason for this nomenclature will get clear later on-- read
\begin{eqnarray}
\hat{\chi}^{(n)}_{\perp}(k)&=&\frac{1}{2}\int\frac{\textrm{d}^{3}k'}{(2\pi)^{3}}g(|\vec{k}'-\vec{k}|)
\Bigl[(1+\cos^{2}{\theta})\nonumber\\&\times&\hat{\chi}^{(n-1)}_{\perp}(k')\Sigma_{\perp}(k')G^{(0)}_{\perp}(k')\nonumber\\
&+&\sin^{2}{\theta}\:\hat{\chi}^{(n-1)}_{\parallel}(k')\Sigma_{\parallel}(k')G^{(0)}_{\parallel}(k')\Bigr],\label{chinper}\\
\hat{\chi}^{(n)}_{\parallel}(k)&=&\int\frac{\textrm{d}^{3}k'}{(2\pi)^{3}}g(|\vec{k}'-\vec{k}|)
\Bigl[\cos^{2}{\theta}\nonumber\\&\times&\hat{\chi}^{(n-1)}_{\parallel}(k')\Sigma_{\parallel}(k')G^{(0)}_{\parallel}(k')
\nonumber\\&+&\sin^{2}{\theta}\:\hat{\chi}^{(n-1)}_{\perp}(k')\Sigma_{\perp}(k')G^{(0)}_{\perp}(k')\Bigr],\label{chinpar}
\end{eqnarray}
for $n\geq1$ and $\hat{\chi}^{(0)}_{\perp,\parallel}(k)\equiv-\tilde{k}^{2}\Sigma_{\perp,\parallel}(k)^{-1}$ for n=0.
In Eqs.(\ref{chinper},\ref{chinpar}), $\Sigma_{\perp,\parallel}(k')$  can be factored out of the integral as $\Sigma_{\perp,\parallel}(k)$ in application of the approximation in Eq.(\ref{laprox}). We notice that $\hat{\chi}^{(1)}_{\perp,\parallel}(k)$ relates to the cavity factors firstly defined in \cite{MeI},
\begin{eqnarray}
C_{\perp}(k)&=&\frac{1}{2}\int\textrm{d}^{3}r\:e^{i\vec{k}\cdot\vec{r}}h_{C}(r)\textrm{Tr}
\{\bar{G}^{(0)}(r)[\bar{I}-\hat{k}\otimes\hat{k}]\}\nonumber\\&=&
\frac{1}{2}\int\frac{\textrm{d}^{3}k'}{(2\pi)^{3}}
h_{C}(|\vec{k}'-\vec{k}|)\Bigl[G_{\perp}^{(0)}(k')\nonumber\\&+&G_{\perp}^{(0)}(k')\cos^{2}{\theta}\:+\:G_{\parallel}^{(0)}(k')\sin^{2}{\theta}\Bigr],
\label{Xioperp}\\
C_{\parallel}(k)&=&\int\textrm{d}^{3}r\:e^{i\vec{k}\cdot\vec{r}}h_{C}(r)\textrm{Tr}
\{\bar{G}^{(0)}(r)[\hat{k}\otimes\hat{k}]\}\nonumber\\&=&\int\frac{\textrm{d}^{3}k'}{(2\pi)^{3}}
h_{C}(|\vec{k}'-\vec{k}|)\nonumber\\&\times&\Bigl[G_{\parallel}^{(0)}(k')\cos^{2}{\theta}\:+\:G_{\perp}^{(0)}(k')\sin^{2}{\theta}\Bigr],
\label{Xioparall}
\end{eqnarray}
 by $\hat{\chi}^{(1)}_{\perp,\parallel}(k)=-\tilde{k}^2[ G^{(0)}_{\perp,\parallel}(k)+C_{\perp,\parallel}(k)]$ in a homogeneous host medium. Also, Eqs.(\ref{gammabnper},\ref{gammabnpar}) resemble the formulae of the $\gamma$-factors for the virtual cavity scenario --Eqs.(\ref{LDOSIper},\ref{LDOSIparal}). That is, if we define the \emph{total pseudo-susceptibility} as
$\hat{\chi}_{\perp,\parallel}(k)\equiv\sum_{n=0}\hat{\chi}_{\perp,\parallel}^{(n)}(k)$
we can write
\begin{eqnarray}
2\gamma_{\perp}&=&2\int\frac{\textrm{d}^{3}k}{(2\pi)^{3}}\frac{\hat{\chi}_{\perp}(k)}{\hat{\chi}_{\perp}^{(0)}(k)}\:G_{\perp}^{(0)}(k),\label{gammaperp}\\
\gamma_{\parallel}&=&\int\frac{\textrm{d}^{3}k}{(2\pi)^{3}}\frac{\hat{\chi}_{\parallel}(k)}{\hat{\chi}_{\parallel}^{(0)}(k)}\:G_{\parallel}^{(0)}(k).\label{gammapara}
\end{eqnarray}
However, the above equations are not yet quite similar to those for the virtual cavity scenario. In particular, the propagator in the integrand of Eqs.(\ref{gammaperp},\ref{gammapara}) is that of free-space whereas it is the bulk propagator for the virtual cavity scenario. The reason being that $\hat{\chi}_{\perp,\parallel}(k)$ contain both 1PI and non-1PI processes. We can decompose  $\hat{\chi}_{\perp,\parallel}(k)$  in 1PI and non-1PI (N1PI) pieces, $\hat{\chi}_{\perp,\parallel}(k)=\hat{\chi}^{1PI}_{\perp,\parallel}(k)+\hat{\chi}^{N1PI}_{\perp,\parallel}(k)$
according to the following decomposition in partial pseudo-susceptibility functions:
\begin{eqnarray}
\hat{\chi}^{N1PI(n)}_{\perp,\parallel}(k)&=&\hat{\chi}^{(n-1)}_{\perp,\parallel}(k)\Sigma_{\perp,\parallel}(k)
G^{(0)}_{\perp}(k),\quad n\geq1,\nonumber\\
\hat{\chi}^{1PI(n)}_{\perp,\parallel}(k)&=&
\hat{\chi}^{(n)}_{\perp,\parallel}(k)-\hat{\chi}^{N1PI(n)}_{\perp,\parallel}(k),\quad n\geq1,\nonumber\\
\hat{\chi}^{1PI(0)}_{\perp,\parallel}(k)&=&\hat{\chi}^{(0)}_{\perp,\parallel}(k),\quad
\hat{\chi}^{N1PI(0)}_{\perp,\parallel}(k)=0.
\end{eqnarray}
Using this decomposition, we can write Eqs.(\ref{gammaperp},\ref{gammapara}) in the form,
\begin{eqnarray}
2\gamma_{\perp}&=&2\int\frac{\textrm{d}^{3}k}{(2\pi)^{3}}\frac{\hat{\chi}^{1PI}_{\perp}(k)}
{\hat{\chi}_{\perp}^{(0)}(k)}\:\tilde{G}_{\perp}(k),\label{gammaperp2}\\
\gamma_{\parallel}&=&\int\frac{\textrm{d}^{3}k}{(2\pi)^{3}}\frac{\hat{\chi}^{1PI}_{\parallel}(k)}
{\hat{\chi}_{\parallel}^{(0)}(k)}\:\tilde{G}_{\parallel}(k).\label{gammapara2}
\end{eqnarray}
where
\begin{equation}\label{pseudoG}
\tilde{G}_{\perp,\parallel}(k)\equiv G_{\perp,\parallel}^{(0)}(k)+G^{(0)}_{\perp,\parallel}(k)
\frac{\hat{\chi}^{N1PI}_{\perp\parallel}(k)}{\hat{\chi}_{\perp\parallel}^{1PI}(k)}\Sigma_{\perp,\parallel}(k)
G^{(0)}_{\perp,\parallel}(k).
\end{equation}
From this equation we can identify a $T$-matrix for the emission process,
\begin{equation}\label{Tmatrix}
\hat{T}_{\perp,\parallel}(k)=\frac{\hat{\chi}^{N1PI}_{\perp\parallel}(k)}{\hat{\chi}_{\perp\parallel}^{1PI}(k)}
\Sigma_{\perp,\parallel}(k).
\end{equation}
\indent Applying the same reasoning as that  in \cite{MeII,MeIII}, we decompose the above function in propagating (P) and non-propagating (NP) components --in absence of longitudinal propagating modes. That is,
\begin{eqnarray}
2\gamma^{P}&=&2\int\frac{\textrm{d}^{3}k}{(2\pi)^{3}}\Re{\{\frac{\hat{\chi}^{1PI}_{\perp}(k)}
{\hat{\chi}_{\perp}^{(0)}(k)}\}}\:\Im{\{\tilde{G}_{\perp}(k)\}},\label{gammaPro}\\
\gamma^{NP}&=&\gamma_{\parallel}+2\int\frac{\textrm{d}^{3}k}{(2\pi)^{3}}\Im{\{\frac{\hat{\chi}^{1PI}_{\perp}(k)}
{\hat{\chi}_{\perp}^{(0)}(k)}\}}\:\Re{\{\tilde{G}_{\perp}(k)\}}.\label{gammaNoPro}
\end{eqnarray}
\indent The above equations are suitable for any real cavity satisfying $R\gg\xi$. A generic case will be treated numerically in a separate paper. Here, in order to make contact with previous results \cite{MeIII,deVriesPRL}, we will treat a simpler realistic scenario. That is, we will compute the spontaneous emission of a two-level atom which seats within a small molecule embedded in a continuous host medium. By \emph{small} molecule we mean that the molecule radius $R_{1}$ is such that $R_{1}\ll\lambda_{res}$. By \emph{continuous} medium we mean that $R_{1}\gg\xi$ being $\xi$ the correlation length of the host medium constituents as in \cite{MeIII} --see Fig.\ref{FigIV03}.\\
\begin{figure}[h]
\includegraphics[height=11.5cm,width=17.cm,clip]{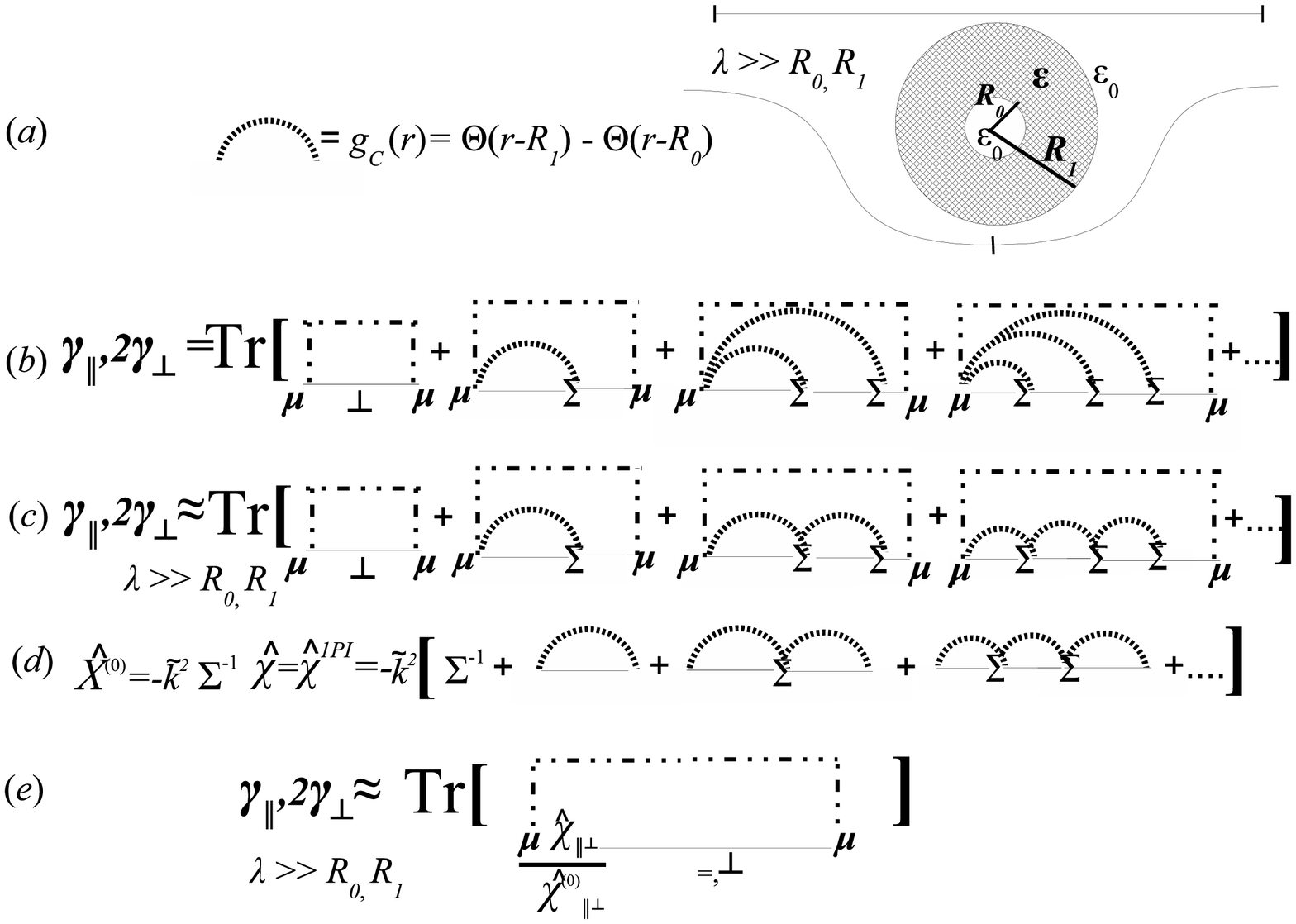}
\caption{($a$) Schematic picture of a point emitter seated at the center of a small molecule of radius $R_{1}$ in free space. $R_{0}$ is the radius of the emitter cavity in the molecular medium. ($b$) Diagrammatic representation of the actual expansion of the $\gamma$-factors according to Eqs.(\ref{seriesa},\ref{gammaanper},\ref{gammaanpar}). ($c$) Diagrammatic representation of the $\gamma$-factors expansion in the small molecule limit. ($d$) Diagrammatic representation of Eq.(\ref{Xi1pispa1}) in the same approximation. ($e$) Diagrammatic representation of the summation of the above series in the small molecule approximation in free space.}\label{FigIV02}
\end{figure}
\indent Consider first the molecule is in vacuum as depicted in Fig.\ref{FigIV02}($a$). The molecular medium surrounding the emitter atom can be thought of as a continuous effective medium of permitivity $\epsilon_{1}$ within $R_{0}<r<R_{1}$. Hence, the cavity correlation function reads
\begin{equation}\label{lagcmol}
g_{C}(r)=\Theta(r-R_{1})-\Theta(r-R_{0}).
\end{equation}
Thus, $g_{C}$ is totally irreducible and so are $\hat{\chi}^{(1)}_{\perp,\parallel}(k)$ and $\hat{\chi}_{\perp,\parallel}(k)$. As a matter of fact,
\begin{equation}\label{xi1}
\hat{\chi}^{(1)}_{\perp,\parallel}(k)=-\tilde{k}^{2}[C_{\perp,\parallel}^{R_{0}}-C_{\perp,\parallel}^{R_{1}}],
\end{equation}
where the superscripts denote the radii of the cavities and Eqs.(\ref{Xioperp},\ref{Xioparall}) are used.
Further on, in the small molecule limit $R_{1}\ll\lambda_{res}$, the wave cannot resolve the molecule structure at all, so that the field is roughly uniform in $R_{0}<r<R_{1}$ and we can write --see Figs.\ref{FigIV02}($c,d$)--
\begin{equation}
\hat{\chi}_{\perp,\parallel}(k)=\hat{\chi}^{(0)}+\hat{\chi}_{\perp,\parallel}^{(1)}(k)
[1-\hat{\chi}^{(1)}_{\perp,\parallel}(k)/\hat{\chi}^{(0)}]^{-1}.\label{Xi1pispa1}
\end{equation}
That is, we can interpret this result as if the wave visited infinite times the molecule before propagating  throughout free space and got back. In this case --and only in this case, the $\gamma$-factors reduce to those formulae in Eqs.(\ref{gammaperp},\ref{gammapara}) with $\hat{\chi}_{\perp,\parallel}(k)$ given by Eqs.(\ref{xi1},\ref{Xi1pispa1}) and $\hat{\chi}^(0)=1/(\epsilon_{1}-1)$. From those equations it is easy to verify that the $\gamma$-factors can be expanded in powers of $(\epsilon_{1}-1)$. $2\gamma_{\perp}(\tilde{k})$ can be computed analytically in closed form. We give below the leading order terms up to order $(\epsilon_{1}-1)^{2}$,
\begin{eqnarray}
2\gamma_{\perp}(\tilde{k})&\simeq&-\frac{i}{2\pi}\tilde{k}+2\Re{\{\gamma_{\perp}^{(0)}\}}\nonumber\\
&-&\frac{i}{2\pi}\tilde{k}\:(\epsilon_{1}-1)\Bigl[\frac{i}{2\tilde{k}}\Bigl(\frac{1}{R_{1}}-\frac{1}{R_{0}}\Bigr)
+\frac{11\tilde{k}^{2}}{30}(R_{1}^{2}-R^{2}_{0})\nonumber\\&+&\frac{2i\tilde{k}^{3}}{9}
(R_{1}^{3}-R^{3}_{0})-\frac{23\tilde{k}^{4}}{210}(R_{1}^{4}-R^{4}_{0})\Bigr]\nonumber\\
&-&\frac{i}{2\pi}\tilde{k}\:(\epsilon_{1}-1)^{2}\frac{121\tilde{k}^{4}}{900}(R_{1}^{2}-R^{2}_{0})^{2}.
\end{eqnarray}
It is remarkable that, in the small molecule limit, the Lorentz-Lorentz (LL) local field factors \cite{Lorentz} cancel out with respect to the computation of \cite{MeIII} for $R_{1}\rightarrow\infty$ and in accordance with the numerical computation of \cite{Luis}. Also, for a given power $n$ of $(\epsilon_{1}-1)$, the leading term in $(\tilde{k}R)$ is of the order of $2n$ but for $n=1$ which contains a non-propagating contribution which goes as $\sim\frac{1}{\tilde{k}R}$. On the contrary, the series expansion of $\gamma_{\parallel}(\tilde{k})$ contains near field contributions at all order $n$ in $(\epsilon_{1}-1)$. That is, we can write the series for $\gamma_{\parallel}(\tilde{k})$ as
\begin{eqnarray}
\gamma_{\parallel}(\tilde{k})=\gamma^{(0)}_{\parallel}&+&\frac{\tilde{k}}{2\pi}\sum_{n=1}(\epsilon_{1}-1)^{n}
\Bigl[\frac{A_{3}^{(n)}}{(\tilde{k}R_{1})^{3}}+\frac{B_{3}^{(n)}}{(\tilde{k}R_{0})^{3}}
\nonumber\\&+&\frac{A_{1}^{(n)}}{\tilde{k}R_{1}}+\frac{B_{1}^{(n)}}{\tilde{k}R_{0}}\Bigr],
\end{eqnarray}
where $A^{(n)}_{1,3}$ and $B^{(n)}_{1,3}$ are real numbers of order 1. In particular, up to order $n=2$,
$A^{(1)}_{3}=1, B^{(1)}_{3}=-1,A^{(1)}_{1}=1/2, B^{(1)}_{1}=-1/2,A^{(2)}_{3}=-2/3, B^{(2)}_{3}=2,A^{(2)}_{1}=0, B^{(2)}_{1}=4/3$.\\
\indent Let us introduce next the molecule of radius $R_{1}$ inside a continuous medium of permitivity $\epsilon_{2}$ as shown in Fig.\ref{FigIV03}. This way, the scenario is similar to that addressed in \cite{MeIII} in which the emitter seats at the center of a small real cavity drilled in a continuous medium. The difference here being that the cavity itself is a molecule which hosts the atomic emitter. The series of diagrams is that of  Fig.\ref{FigIV03}. It is analogous to that in \cite{MeIII} but replacing the factors
$C^{R_{1}}_{\perp,\parallel}(k)+G^{(0)}_{\perp,\parallel}(k)$  with $\frac{\hat{\chi}_{\perp,\parallel}(k)}
{\hat{\chi}^{(0)}}[C^{R_{1}}_{\perp,\parallel}(k)+G^{(0)}_{\perp,\parallel}(k)]$. We will restrict ourselves to the computation of the propagating decay rate, $2\Gamma^{P}(\tilde{k})=\frac{-2\pi}{\tilde{k}}\Gamma_{0}\Im{\{2\gamma_{\perp}^{P}\}}(\tilde{k})$, where $\tilde{k}$ is the resonant frequency in the medium satisfying Eq.(\ref{kres}) and $\Gamma_{0}$ is the decay rate of the atom in absolute vacuum. $\Im{\{2\gamma_{\perp}^{P}\}}(\tilde{k})$ reads,
\begin{eqnarray}
\label{firstp}\Im{\{2\gamma_{\perp}^{P}(\tilde{k})\}}=2\Re{\{\frac{\hat{\chi}_{\perp}}
{\hat{\chi}^{(0)}}(\tilde{k})\}}\int\Im{\{G_{\perp}(k)\}}\frac{\textrm{d}^{3}k}{(2\pi)^3}\\
-2\tilde{k}^{2}\Re{\{\frac{\hat{\chi}_{\perp}}
{\hat{\chi}^{(0)}}(\epsilon_{2}-1)
C^{R_{1}}_{\perp}(\tilde{k})\}}\int\Im{\{G^{(0)}_{\perp}(k)\}}\frac{\textrm{d}^{3}k}{(2\pi)^3}\nonumber\\
+2\tilde{k}^{4}\Re{\{[\frac{\hat{\chi}_{\perp}}
{\hat{\chi}^{(0)}}(\epsilon_{2}-1)
C^{R_{1}}_{\perp}(\tilde{k})]^{2}\}}\int\Im{\{G_{\perp}(k)\}}\frac{\textrm{d}^{3}k}{(2\pi)^3}\nonumber\\
+4\tilde{k}^{4}\Re{\{\frac{\hat{\chi}_{\perp}}
{\hat{\chi}^{(0)}}(\epsilon_{2}-1)
C^{R_{1}}_{\perp}(\tilde{k})\}}\int\Im{\{G_{\perp}^{(0)}(\epsilon_{2}-1)
G_{\perp}(k)\}}\frac{\textrm{d}^{3}k}{(2\pi)^3},\nonumber
\end{eqnarray}
where $G_{\perp}(k)=\frac{1}{\epsilon_{2}\tilde{k}^2-k^2}$ is the transverse component of the bulk propagator in the continuous medium. For the sake of simplicity, let us assume $\epsilon_{1},\epsilon_{2}$ are real in first approximation. In such a case, we obtain
\begin{eqnarray}\label{lagamma}
2\Gamma_{\perp}^{P}(\tilde{k})&=&2\Gamma^{P}_{2}+\frac{11}{30}\Gamma_{0}\Bigl[(\epsilon_{1}-1)\sqrt{\epsilon_{2}}
[(\tilde{k}R_{1})^{2}-(\tilde{k}R_{0})^{2}]\nonumber\\&-&\frac{1}{3}(\epsilon_{2}-1)[(\tilde{k}R_{1})^{2}(4-\epsilon_{1})+
(\epsilon_{1}-1)(\tilde{k}R_{0})^{2}]\nonumber\\&\times&[1+\frac{2}{3}\sqrt{\epsilon_{2}}-2(\epsilon_{2}-1)(\sqrt{\epsilon_{2}}-1)]\Bigr].
\end{eqnarray}
Where $2\Gamma^{P}_{2}$ is the formula found in \cite{MeIII} when  the finite size of both the molecule and the emitter cavity are discarded. It does contain the LL local field factors \cite{Lorentz,Bullough},
\begin{equation}
2\Gamma_{2}^{P}=\Gamma_{0}\Bigl[\Bigl(\frac{\epsilon_{2}+2}{3}\Bigr)^{2}\sqrt{\epsilon_{2}}-\frac{1}{3}(\epsilon_{2}-1)\Bigr].
\end{equation}
The additional terms account for finite size corrections at leading order in $\tilde{k}R$. As argued above, the expansion of $\hat{\chi}_{\perp}/\hat{\chi}^{(0)}$ in powers of $(\epsilon_{1}-1)$ yields terms $\sim(\epsilon_{1}-1)^{n}(\tilde{k}R)^{2n}$ at leading order in $(\tilde{k}R)$. Therefore, in the range of frequencies where the small molecule limit is valid, strong deviations are expected with respect to Eq.(\ref{lagamma}) close to resonances of the permitivity  of the intramolecular medium. That is, for frequencies $c\tilde{k}$ such that $\tilde{k}R_{1}\ll1$ but $|(\epsilon_{1}-1)(\tilde{k}R_{1})^{2}|\gtrsim1$.\\
\indent In summary, we have derived generic expressions for the $\gamma$ radiative factors which dress up the bare polarizability of a point emitter which seats within a macroscopic real cavity ($R\gg\xi$) drilled in a random medium --Eqs.(\ref{gammaperp2},\ref{gammapara2}) together with Eq.(\ref{pseudoG}). To that aim we define pseudo-susceptibility functions, $\hat{\chi}_{\perp,\parallel}(k)$, whose partial components satisfy the recurrent relations in Eqs.(\ref{chinper},\ref{chinpar}). The formulae so obtained are analogous to those found in \cite{MeI,MeII} in a virtual cavity scenario. Propagating and non-propagating emission rates are identified --Eqs.(\ref{gammaPro},\ref{gammaNoPro}). For the case of a single atom which seats within a small molecule embedded in a continuous medium an expression for the pseudo-susceptibility is found in closed form --Eqs.(\ref{xi1},\ref{Xi1pispa1})-- in terms of the cavity factors --Eqs.(\ref{Xioperp},\ref{Xioparall}). An expression for the propagating emission rate is found as a function of the radii of the molecule and the emitter cavity within the molecule and the permitivities of the intramolecular medium and the host medium --Eq.(\ref{lagamma}).

\end{document}